\documentclass[]{spie}  

\usepackage{amsmath,amsfonts,amssymb}
\usepackage{graphicx}
\usepackage[colorlinks=true, allcolors=blue]{hyperref}
\usepackage{makecell}

\title{Threat determination for radiation detection from the Remote Sensing Laboratory}

\author[a]{William P. Ford}
\author[a]{Emma Hague}
\author[a]{Tom McCullough}
\author[a]{Eric Moore}
\author[a]{Johanna Turk}
\affil[a]{Remote Sensing Laboratory, 1783 Arnold Ave., Joint Base Andrews 20762, US}

\authorinfo{Further author information: E-mail: fordwp@nv.doe.gov, Telephone: 1 (301) 817-3358}

\begin{document} 
\maketitle

\begin{abstract}
The ability to search for radiation sources is of interest to the Homeland Security community. The hope is to find any radiation sources which may pose a reasonable chance for harm in a terrorist act. The best chance of success for search operations generally comes with fielding as many detection systems as possible. In doing this, the hoped for encounter with the threat source will inevitably be buried in an even larger number of encounters with non-threatening radiation sources commonly used for many medical and industrial use. The problem then becomes effectively filtering the non-threatening sources, and presenting the human-in-the-loop with a modest list of potential threats. Our approach is to field a collection of detection systems which utilize soft-sensing algorithms for the purpose of discriminating potential threat and non-threat objects, based on a variety of machine learning techniques.
\end{abstract}


\section{INTRODUCTION}
\label{sec:intro}  

Radiological weapons pose a serious concern. In order to mitigate this concern 
Preventative Radiological Nuclear Detection (PRND) is performed before and during major public events.
The traditional method of performing PRND consists of deploying personnel equipped with radiological detectors around the event.
This approach is limited by the amount of personel available, their expertise in using the detectors, and interpreting the data.
False alarms are common, as confuser sources e.g. medical isotopes and industrial gauges, are numerous. 
In addition the radiological background varies significantly based on immediate surroundings further complicating threat determination. 
In order to deal with false alarms, spectroscopists are utilized to review data and determine whether an anomaly is a threat.
Therefore PRND is fundamentally limited by the number of trained personnel, both in fielding detectors, and in interpreting the results.
Our approach is to employ a vast network of static sensors, monitored by neural networks allowing a single spectroscopist to monitor orders of magnitude more sensors than is traditionally possible. With such a network augmenting classical PRND, we can have an increased confidence in detecting threats.
This paper discusses our implementation of machine learning techniques for isotope identification of gamma ray spectra.
\section{Framework}\label{sec:Framework}
\begin{figure} [ht]  
	\begin{center}
		\begin{tabular}{c} 
			\includegraphics[height=10cm]{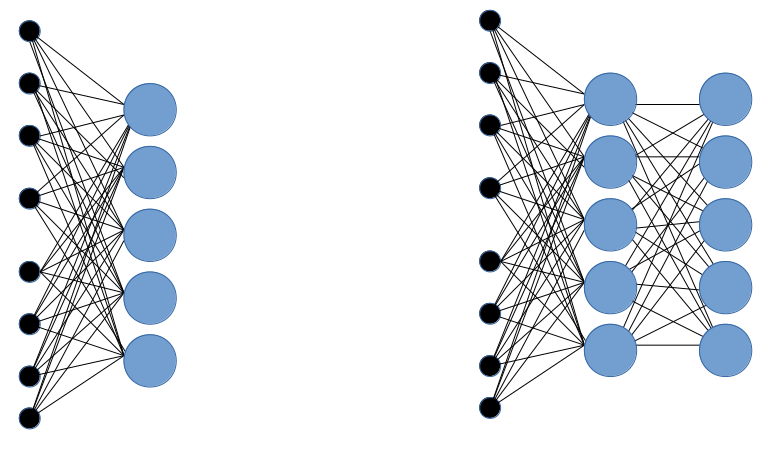}
		\end{tabular}
	\end{center}
	\caption[example] 
	{ \label{fig:networks} 
		Representations of the network architectures employed. The network on the left represents a linear model and the network on the right is a single hidden layer neural network (NN). The activation function on the NN hidden layer was $tanh$. The number of weights and neurons depicted are representative, not literal, and were varied throughout the study. }
\end{figure}

For this work we employed two simple networks depicted in Fig. \ref{fig:networks}. The single layer network (left in Fig. \ref{fig:networks}) can be expressed as
\begin{equation} \label{eq:single_layer}
\hat{\mathbf{y}} = \left( \mathbf{W}\cdot \mathbf{x} + \mathbf{b}  \right),
\end{equation}
where $\mathbf{x}$ is the input vector, $\mathbf{W}$ is a matrix of the weights, 
$\mathbf{b}$ are the bias', and $\mathbf{\hat{y}}$ is a softmax normalized output vector representing the various classes. 
For this work the input vector is a gamma spectrum of 1024 channels (or rebinned to 256), and the elements of the output vector are the isotopes we are considering. The slightly more complex network shown on the right in Fig. \ref{fig:networks} can likewise be expressed as
\begin{align}
\hat{\mathbf{y}} &=  \mathbf{W_2}\cdot \mathbf{y_1} + \mathbf{b_2},   \\
\mathbf{y_1} &= \tanh \left( \mathbf{W_1}\cdot \mathbf{x} + \mathbf{b_1}  \right),
\end{align}
where the $\tanh $ function implies elementwise operation.

The networks were implemented and trained using Tensorflow \cite{tensorflow2015-whitepaper}. 
The cost function was chosen to be cross entropy. 
Training was performed using AdamOptimizer. 

For this work we generate data from modeled, or simulated, and measured spectra. 
The first source is simulated spectra using GADRAS \cite{Gadras-Manual,Gadras-Transport}. 
Gadras provides deterministic gamma and neutron transport, and has a variety of detectors characterized. 
For the purposes of this work we focus on $2"$ x $4"$ x $16" $ sodium iodide (NaI) detectors.
This was chosen to be consistent with our detector that we used for our data collects.
The spectra are initially produced at 24 hour long dwells, 
and then are Poisson sampled to appropriate dwell times, 
typically 1 second spectra again corresponding to the sampling frequency of our detectors.

\begin{figure} [ht]  
	\begin{center}
		\begin{tabular}{cc} 
			\includegraphics[height=6cm]{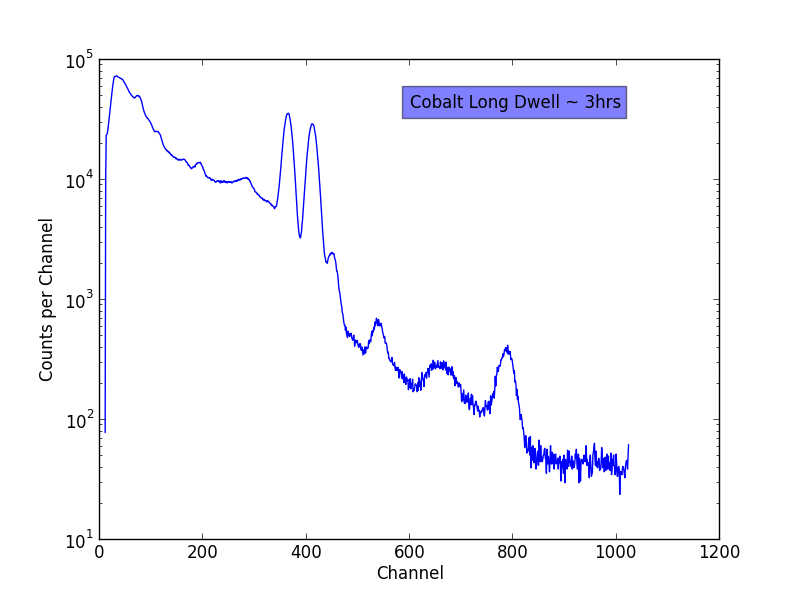}
			\includegraphics[height=6cm]{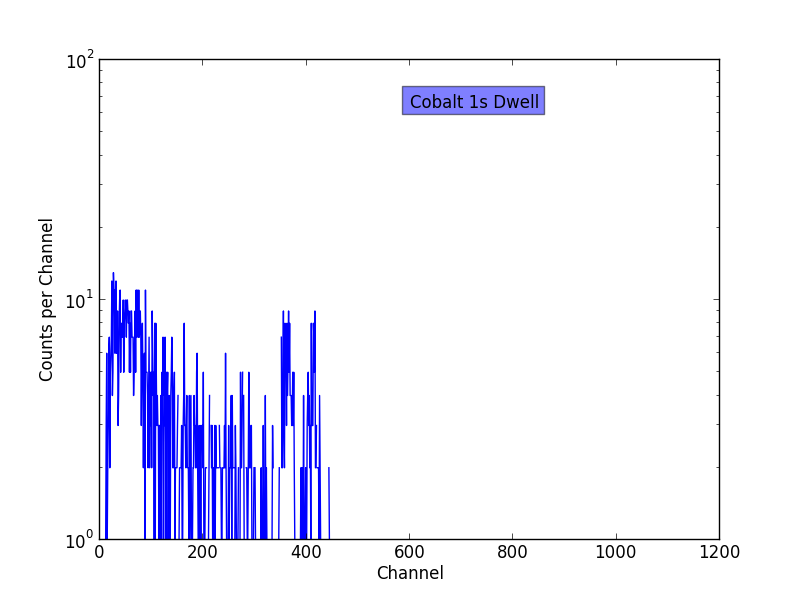}
		\end{tabular}
	\end{center}
	\caption[example] 
	{ \label{fig:examplespectra} 
		Examples of a Cobalt Gamma-ray spectra taken with a $2"x4"x16"$ NaI detector. The one on the left is a long dwell of about 3 hours. 
		The spectrum on the right is a 1 second spectrum, which is a standard sampling frequency, and what we train the machines to identify in this work. }
\end{figure}

The second source of data are measured spectra, which we performed in controlled collects at our lab.
The data were collected using a $2" x 4" x 16" $ sodium iodide (NaI) detector.
Both collected and simulated sources are at a variety of distances and shielding configurations in order to ensure that the machines trained are robust against these variations. These variations are presented in Table \ref{tab:Variations}. We show an example of a long dwell spectrum and a 1 second spectrum that we typically train our machines to identify in Fig. \ref{fig:examplespectra} 

\begin{table}[ht]
	\caption{Variations of sources, distances, and shieldings for data.} 
	\label{tab:Variations}
	\begin{center}       
		\begin{tabular}{|l|l|l|l|}
			\hline
			\rule[-1ex]{0pt}{3.5ex}            & Isotopes & Distances (m) & Shiedings  \\
			\hline
			\rule[-1ex]{0pt}{3.5ex} Simulated  & Cesium, Cobalt, Barium, Selenium, Iridium & 10,11,12,...,20 & \makecell{Bare, Concrete, Steel, \\ Depleted Uranium}      \\
			\hline
			\rule[-1ex]{0pt}{3.5ex} Measured   & Cesium, Cobalt, Barium & ~0-10 & Bare, Steel  \\
			\hline
		\end{tabular}
	\end{center}
\end{table}

\section{Results}\label{sec:Results}
\begin{figure} [ht]  
	\begin{center}
		\begin{tabular}{cc} 
			\includegraphics[height=8cm]{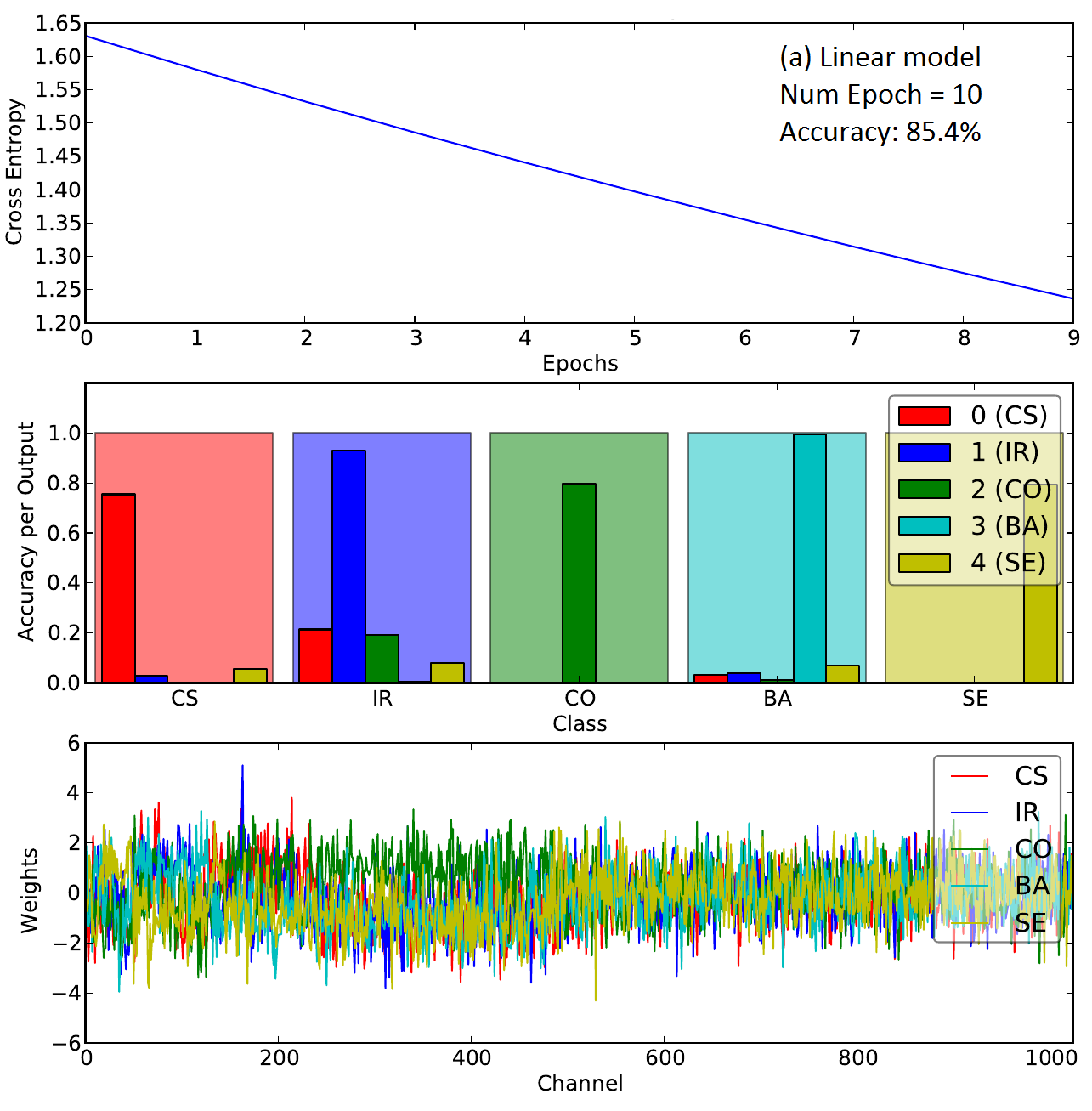}
			\includegraphics[height=8cm]{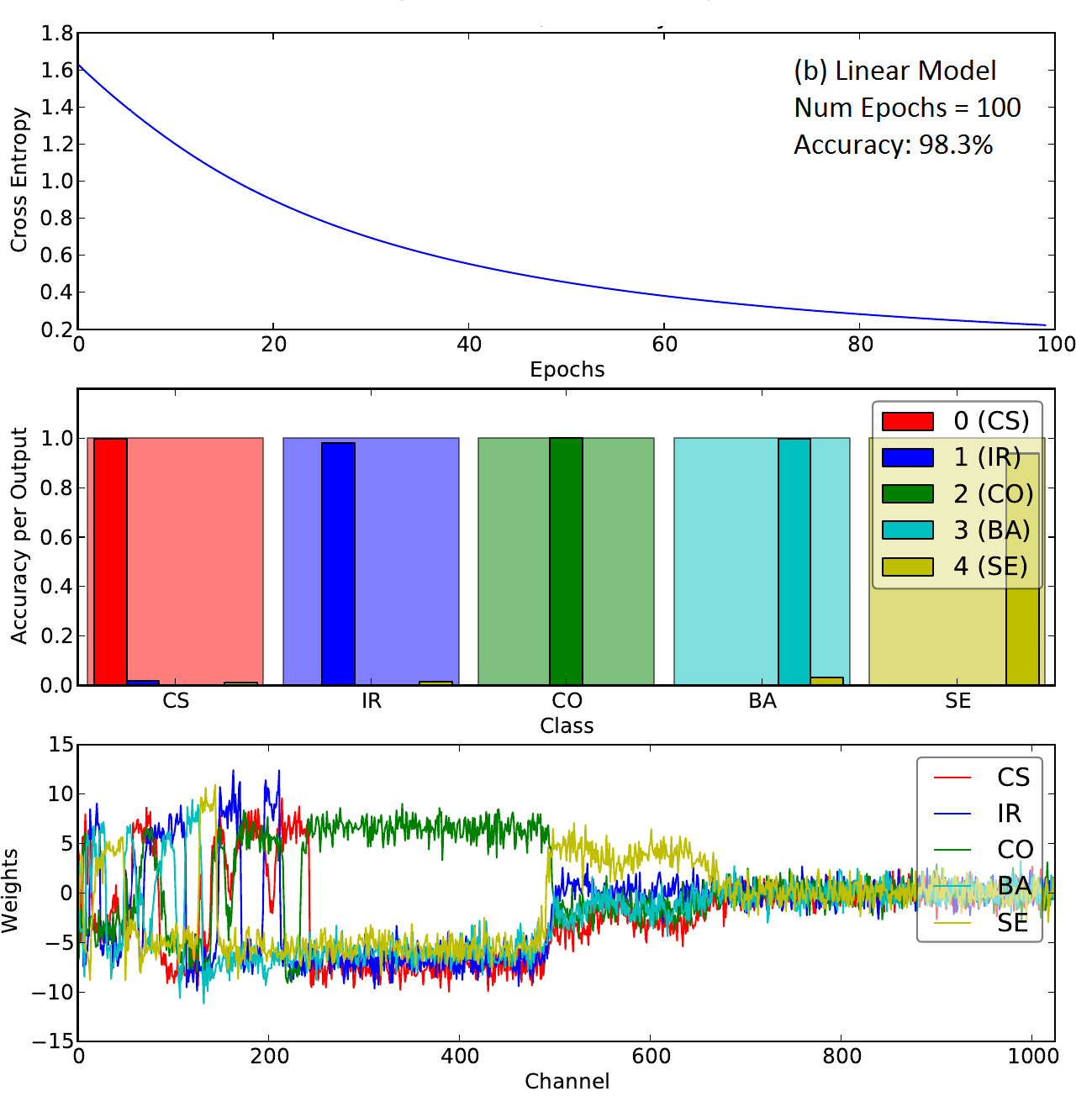}
		\end{tabular}
	\end{center}
	\caption[example] 
	{ \label{fig:Ensemble_Study} 
		Convergence example for training to the simple linear model. The graphs from top to bottom correspond to the cost function, cross entropy, then we show the accuracy for each class, and the bottom graph shows the values of the weights. The plots on the left are after training to ten epochs while the graphs on the right correspond to 100 epochs. Note the appearance of spectral features in the weights in the right after the training has converged. }
\end{figure}

In Fig. \ref{fig:Ensemble_Study} results are shown for training to simulated data and testing against a Poisson sampled ensemble. 
We experimented with training to both the time asymptotic spectra as well as a subset of the ensemble. Preliminary results suggest that it is actually better to train to a subset of the ensemble, however, further studies are needed. The results shown are from training to time asymptotic spectra.
The top figures represent the cost function, Cross Entropy, after training to ten epochs (left) vs. training to 100 epochs (right). 
Also shown is the overall accuracy with regards to the testing data.
the middle plots represent the accuracy per output, i.e. how often the machine classifies the appropriate isotope.
The bottom plots are of the weights, each row of the weight matrix is a different colored line on the plot.
These results illustrate convergence of the training. Notice in the plots on the left the much lower accuracy, both overall and per output. Particularly interesting are the plots of the weights. As the training converges one starts to see spectral features apparent in the weights. Note also that at higher energies (Channels $> 600$) there are no features as these spectra have no counts that high and only noise is reflected in the weights.

\begin{figure} [ht]  
	\begin{center}
		\begin{tabular}{cc} 
			\includegraphics[height=16cm]{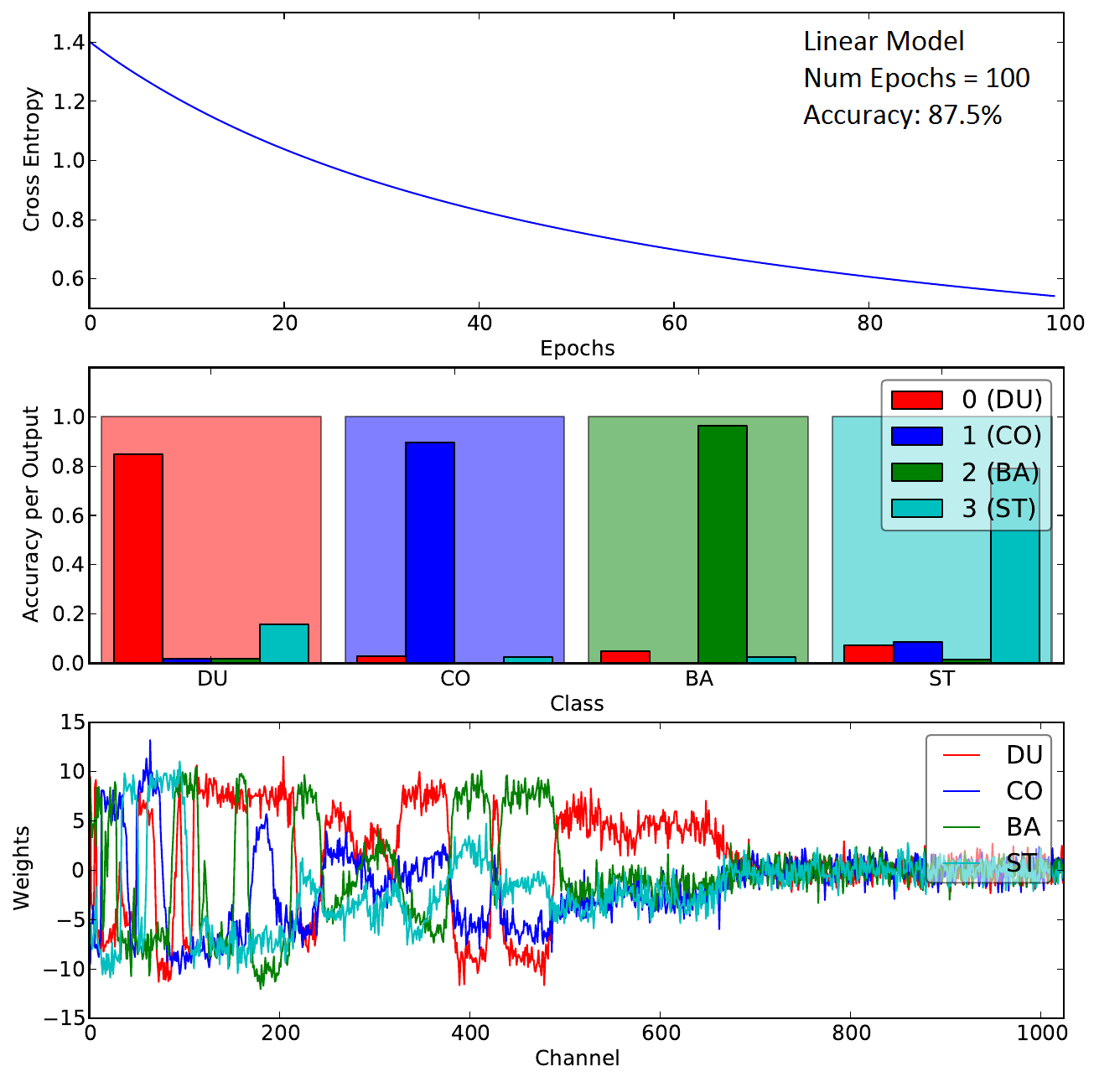}
		\end{tabular}
	\end{center}
	\caption[example] 
	{ \label{fig:Shielding_Study} 
		Results of a single layer machine trained to identify shielding type regardless of what source is behind it. Plots from top to bottom correspond to the cost function, accuracy per output, and values for the weight matrix. These results illustrate robustness of the identifying machine. }
\end{figure}

Besides just trying to classify according to isotope we also wanted to test whether we could identify what type of shieding a source was behind. While this has no direct application it sheds light onto the robustness of the machines. We took the same training set, and trained to identify shielding; these results are shown in Fig. \ref{fig:Shielding_Study}. Note the while some of the features extracted are similar many others are significantly different. In particular the double humps between Channels 500-600 are interesting as they correspond to depleted uranium, and one can see a strong correlation and anti-correlation in the weights. 

\begin{figure} [ht]  
	\begin{center}
		\begin{tabular}{cc} 
			\includegraphics[height=4cm]{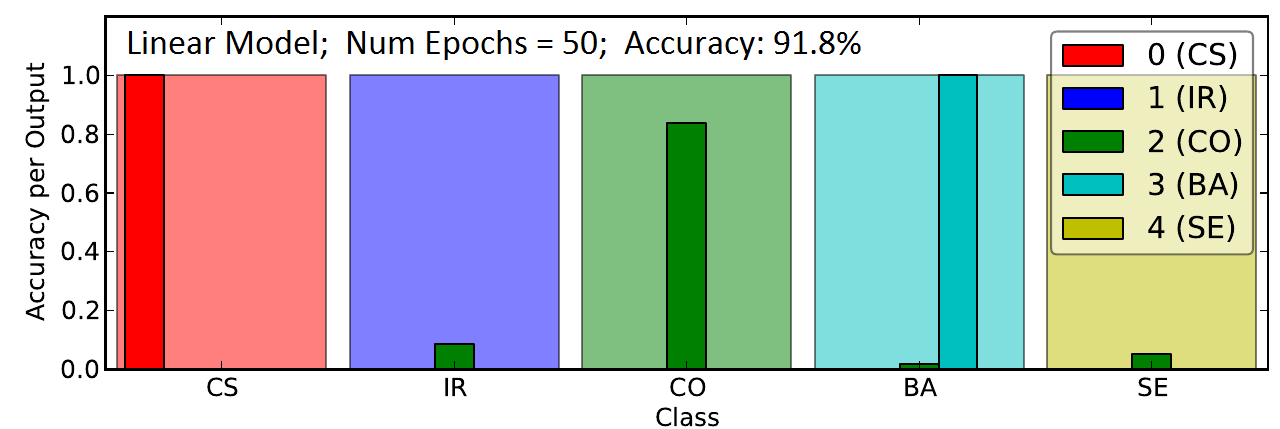}
		\end{tabular}
	\end{center}
	\caption[example] 
	{ \label{fig:Measured_Data} 
		Results of training to simulated data and testing against measured data. }
\end{figure}

Next we turn our attention to testing our machine against real data. For this we utilize the single layer network trained to time asymptotic simulated data, and tested againts data collects taken at our lab. The testing data are 1 second spectra as shown in Fig. \ref{fig:examplespectra}. The results are surprisingly good considering that this machine was not trained to data having any background radiation. The results are shown in Fig. \ref{fig:Measured_Data}   

Each of the above results were also implemented with the hidden layer network with similar results. While we typically had slightly better accuracy the small improvement did not seem to justify the more complicated architecture for these examples. 

\begin{figure} [ht]  
	\begin{center}
		\begin{tabular}{cc} 
			\includegraphics[height=8cm]{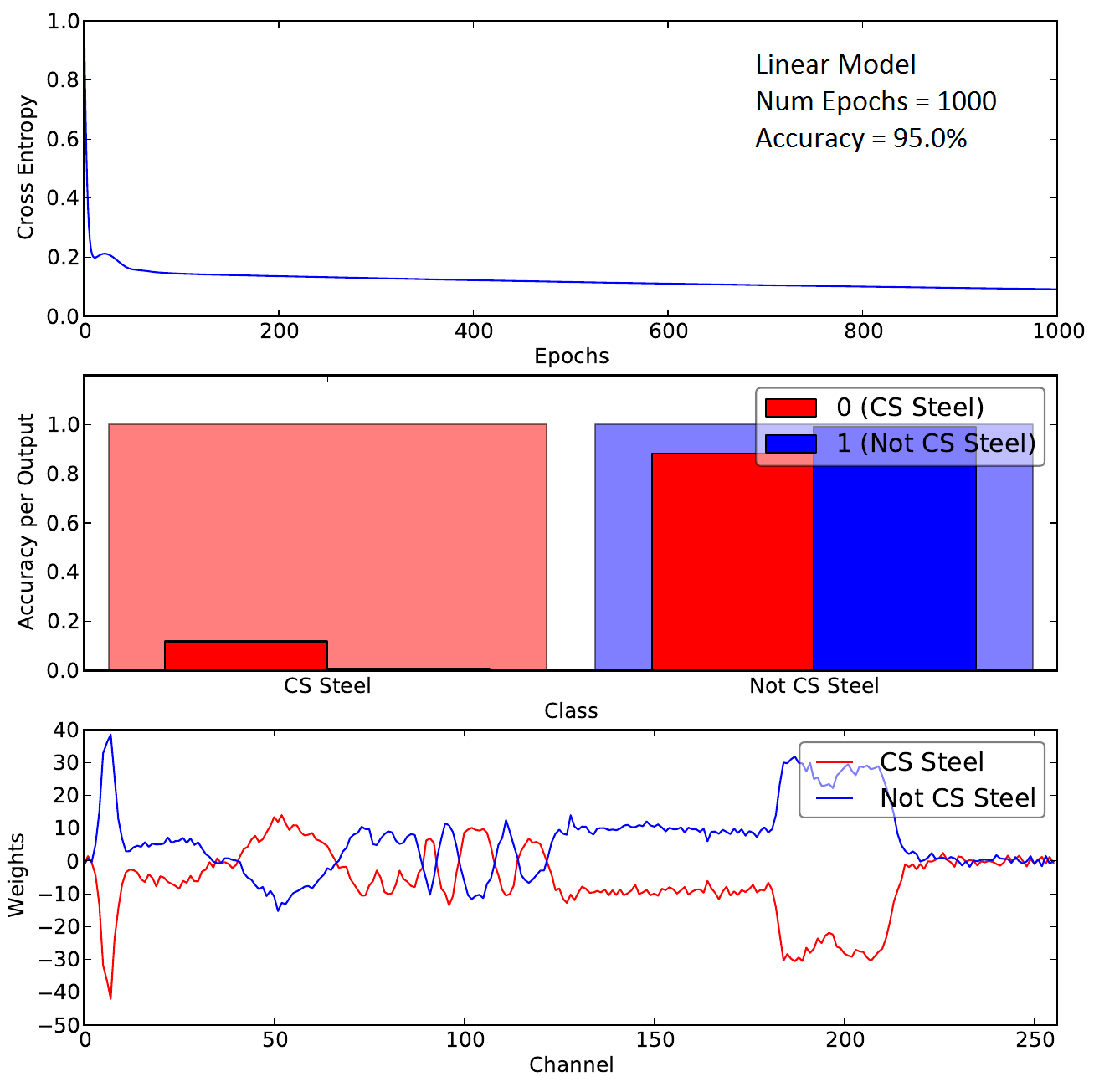}
			\includegraphics[height=8cm]{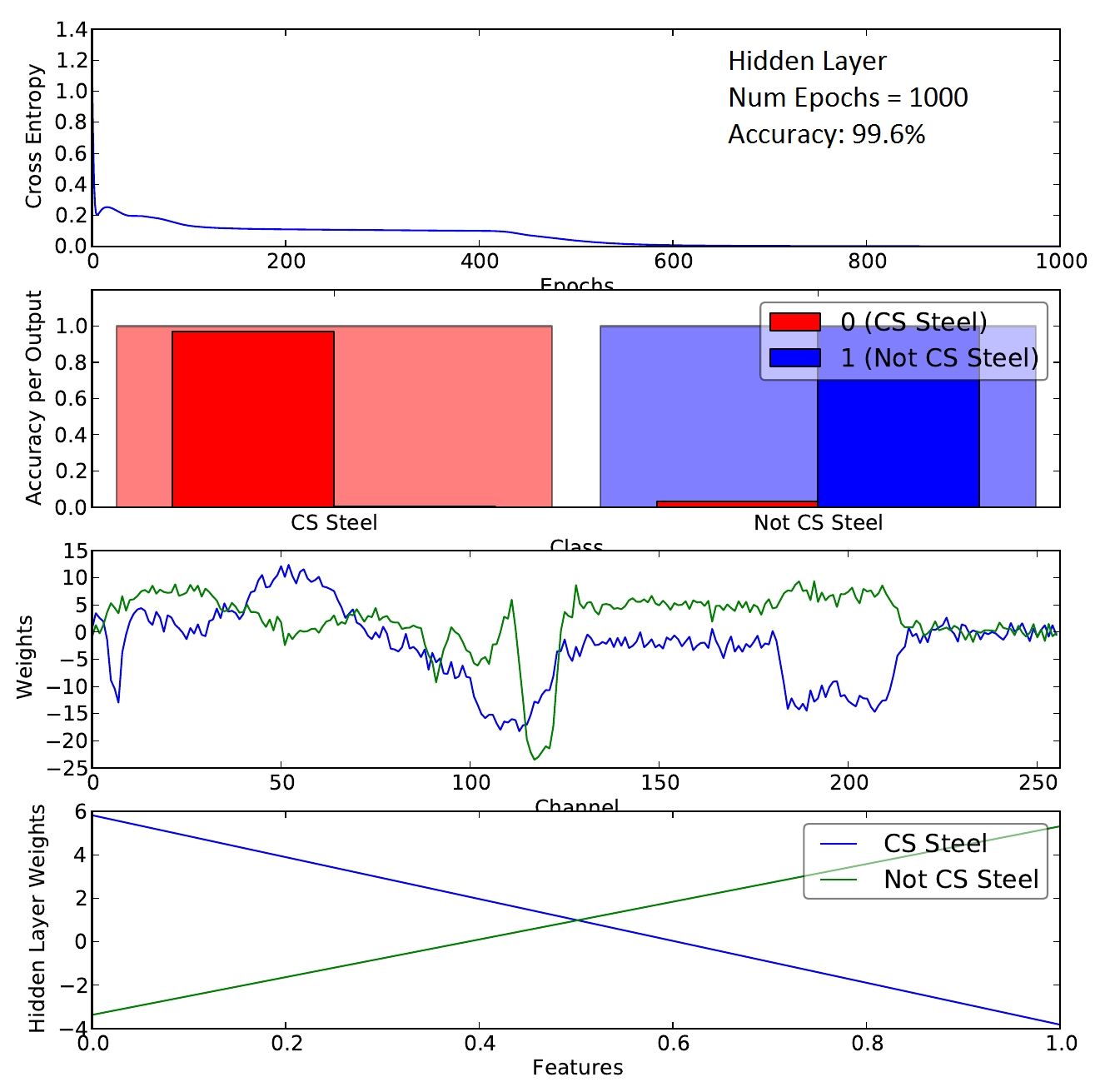}
		\end{tabular}
	\end{center}
	\caption[example] 
	{ \label{fig:Industrial_Results} 
		Results showing surragate industrial gauge classifier. Plots are the same as in Fig. \ref{fig:Ensemble_Study}. Note the failure of the linear model to correctly identify the gauge (accuracy per output $<20\%$) while the hidden layer determines it with almost 100\% accuracy. }
\end{figure}

A typical confuser when performing PRND are industrial gauges. To determine the feasablity of a machine identify such a gauge we defined a surragate industrial gauge as Cesium shielded by steel, and allowed all other sources and configurations to be confusers, including Cesium in the other shielding and distance configurations. We then train the machine to identify Cesium Steel (Industrial Gauge) or Not Cesium Steel (anything else!). The results of this are shown in Fig. \ref{fig:Industrial_Results}. Here we show the results of both the simple linear model and the hidden layer network. In particular we note that while we have good convergence of the linear model it almost completely fails to identify the gauge, i.e. the accuracy for that output is less than $20\%$. The hidden layer network on the other hand is almost completely accurate correctly identifying the gauge. This suggests that there are certain spectral id problems that demand deep learning or at least the nonlinearity that is in the hidden layer network.

\section{Summary and Outlook}\label{sec:Summary}
We have implemented two different machine learning models, a simple linear model and a hidden layer neural network. 
These machines were trained with simulated spectra and were tested against a Poisson sampled ensemble and measured data. 
An important result is that one can train to simulated data and obtain a machine that performs well against measured data. 
We have identified that for many cases a simple linear machine may suffice, but there are cases where non-linear machines vastly outperform the linear model. Future work includes expanding both our simulated and modeled datasets, and with more complex data we expect to need more complex architectures.

\acknowledgments 
This manuscript has been authored by Mission Support and Technical Services, LLC, under Contract No. DE-NA0003624 with the U.S. Department of Energy, National Nuclear Security Administration, Office of Defense Nuclear Nonproliferation Research and Development. The United States Government retains and the publisher, by accepting the article for publication, acknowledges that the United States Government retains a non-exclusive, paid-up, irrevocable, worldwide license to publish or reproduce the published form of this manuscript, or allow others to do so, for United States Government purposes. The U.S. Department of Energy will provide public access to these results of federally sponsored research in accordance with the DOE Public Access Plan (\href{http://energy.gov/downloads/doe-public-access-plan}{http://energy.gov/downloads/doe-public-access-plan}). The views expressed in the article do not necessarily represent the views of the U.S. Department of Energy or the United States Government. DOE/NV/03624--0088

The authors would like to thank Sarah Bender, Andre Butler, Emily Jackson, Lance Mclean, Jessica McNutt, Scott Suchyta, and Julia You 
for their help with this project.    
\bibliography{report} 
\bibliographystyle{spiebib} 

\end{document}